----------
X-Sun-Data-Type: default
X-Sun-Data-Name: bilayer.tex
X-Sun-Content-Lines: 600

\documentstyle[preprint,aps]{revtex}
\begin{document}
\draft
\title{Phase Transitions of the Bilayered Spin-$\bf {S}$ Heisenberg Model and
Its Extension to Fractional Dimensions}
\author{Kwai-Kong Ng$^{1}$, Fu Chun Zhang$^{1,2}$, and Michael Ma$^{1,3}$}
\address{
$^{1}$Department of Physics, University of Cincinnati, Cincinnati, OH 45221\\
$^{2}$Department of Physics, Hong Kong University of Science and Technology\\
Clear Water Bay, Kowloon, Hong Kong\\
$^{3}$Department of Physics, Chinese University of Hong Kong\\
Satin, New Territories, Hong Kong}

\maketitle
\begin{abstract}  \small
We study the ground state and the phase transitions of the bilayered spin-$S$
antiferromagnetic Heisenberg model using the Schwinger boson mean field theory.
The interplane coupling initially stabilizes but eventually destroys the
long-range antiferromagnetic order. The transition to the disordered state is
continuous for small $S$, and first order for large $S$. The latter is
consistent with an argument based on the spin wave theory. The phase diagram
and phase transitions in corresponding model in fractional dimensions are also
discussed.
\end{abstract}


\section{Introduction}
Recently there has been considerable interest in quantum spin liquids,
which are magnetic systems without LRO at low temperature.  While in
general, the ground state of quantum spin systems lack true LRO in 1D, the
ground state of the 2D Heisenberg antiferromagnet exhibits Neel ordering
even for $S = 1/2$, albeit with an sublattice magnetization that is
considerably decreased from its classical value.  Since spin is quantized,
the spin value cannot be decreased beyond $1/2$, hence the model does not
have a spin liquid ground state.  On the other hand, when two planes of
antiferromagnetic spins are coupled together
\cite{Millis,Hida1,Hida2,Hida3,Dagotto,Sandvik1,Sandvik2,Yoshioka,Maekawa}, and
if the interplane
coupling is strong enough, the ground state is easily seen to be one of
valence bond solid of interplane singlets (IVBS).  Thus, there should be a
transition from the LRO Neel state to a spin liquid state as the interplane
coupling is increased.  It has been suggested that  the unusual magnetic
properties of YBCO, with its basic unit of a pair of coupled CuO planes,
may be due to its lying close to this quantum transition \cite{Millis}.

It is of interest to study the nature of this quantum transition.  Within a
non-linear sigma-model (NLSM) description, Haldane \cite{Haldane} has pointed
out that for
a single plane of spins, topological Berry phase terms exist which differ
between half-integer, odd integer, and even integer spins. One way to
understand this is to consider the degeneracy of the valence bond solid
states which maximize the number of resonating plaquettes in each case
(4-fold, 2-fold, and non-degenerate respectively). On the other hand, the
mapping of the two-plane system to the NLSM does not yield a topological
term, which is consistent with the valence bond solid state for two planes
with large interplane vs. intraplane coupling being zero-dimensional like
and non-degenerate.  Since the 2+1 D NLSM has only one phase transition
which is second order, this suggests the same for the 2D quantum Heisenberg
antiferromagnet at T = 0.  However, the NLSM mapping assumes slow variation
on the scale of lattice spacing, and so additional disordered phases and/or
first order transition cannot be ruled out conclusively.

In this paper we investigate the ground state of the 2D bilayered Heisenberg
AF for general $S$ using the Schwinger boson mean field theory
\cite{Arovas,Sarker} with no
additional approximation . Our calculation complements previous
calculations for $S = 1/2$ only and/or using additional approximations, as
well as a calculation using the related Takahasi bosons approach
\cite{Hida1,Hida2,Hida3}.  Our
results show that a first order transition is favored by decreasing quantum
fluctuations, i.e.. increasing $S$.  In particular, in agreement with those
previous works, the transition for $S = 1/2$ is first order.  The critical
$S$ separating first from second order transition equals $0.35$.  While
increasing interplane coupling $J_ {\perp}$ eventually destroys LRO,
ordering is stabilized by small $J_ {\perp}$, and the critical $S$ for no
LRO is shifted from the single plane value of  $0.2$ to $0.13$.  A simple
argument using spin wave theory helps to explain why first order transition
occurs for large $S$.  Since quantum fluctuations increases both with
decreasing $S$ and decreasing $d$, we also study the dimensionality
dependence of the "bilayer" hypercubic system.  For $S = 1/2$, we find that
for $d < 1.86$ , the first order transition is replaced by second order one.
En route, we also calculate the $S$ vs. $d$ phase diagram for $J_ {\perp} =
0$, i.e. the hypercubic Heisenberg AF.  In addition to contradicting the NLSM
description by having the possibility of a first order phase transition,
whenever the transtion is continuous, the Schwinger boson MFT gives an
additional phase transition between two disordered phases for all $S$,
corresponding to a jump in the ratio of short-ranged intraplane to
interplane correlations.  However, since the Schwinger boson order
parameter is not related directly to any physical symmetry breaking, it is
likely this does not constitute a real phase transition but a sharp
cross-over in behavior.

\section{Bilayered Antiferromagnet in 2D}

We begin with a quick review of Schwinger boson mean field theory \cite{Sarker}
as
applied to the translationally invariant nearest neighbor Heisenberg
antiferromagnet on a
bipartite lattice. The Hamiltonian is

\begin{eqnarray}
H=\sum_{\langle ij \rangle} J_{ij}{\bf{S}}_i\cdot{\bf{S}}_j; ~~~~~
J_{ij}>0,~~ \langle
ij \rangle =n.n.
\nonumber
\end{eqnarray}
In the Schwinger boson representation, spin operators in each lattice site
are replaced by spin 1/2 bosons as follows:

\begin{eqnarray}
S^{\dag}_{i} &=& b_{i\uparrow}^{\dag} b_{i\downarrow},~~~~ S^{-}_{i} =
b_{i\downarrow}^{\dag} b_{i\uparrow} \nonumber\\
S^{z}_{i} &=& \frac{1}{2}( b_{i\uparrow}^{\dag} b_{i\uparrow} -
b_{i\downarrow}^{\dag} b_{i\downarrow}), \nonumber
\end{eqnarray}
The number of bosons at each lattice site is subject to the constraint:

\begin{eqnarray}
\sum_{\sigma}b^{\dag}_{i\sigma}b_{i\sigma} = 2 S, \nonumber
\end{eqnarray}
which can be implemented by introducing a Lagrange multiplier on each site.
The Hamiltoninan can now be written as

\begin{eqnarray}
H = -2 \sum_{\langle ij \rangle} J_{ij}\tilde{A}^{\dag}_{ij} \tilde{A}_{ij}
+ \frac{1}{2} NzJS^{2} + \sum_{i}
\lambda_i(\tilde{b}^{\dag}_{i\sigma}\tilde{b}_{i\sigma} -
2S), \nonumber
\end{eqnarray}
where $\tilde{A}^{\dag}_{ij}=\frac{1}{2}\sum_{\sigma}
\tilde{b}^{\dag}_{i\sigma}
\tilde{b}^{\dag}_{j\sigma}$ and
$\tilde{b}_{i\uparrow}=b_{i\downarrow},~\tilde{b}_{i\downarrow}
=-b_{i\uparrow}$ for sites on one sublattice and  $\tilde{b}_{i\sigma} =
b_{i\sigma}$
for sites on the other sublattice.  Physically, the product
$\tilde{A}^{\dag}_{ij} \tilde{A}_{ij}$ acts as the valence bond (singlet)
number operator for sites (i,j).  In the mean field approximation, this
product is decoupled by the Hartree-Fock decomposition.  In addition, the
exact local constraint is relaxed to one  for the average:

\begin{eqnarray}
\langle\sum_{\sigma}b^{\dag}_{i\sigma}b_{i\sigma}\rangle = 2 S \nonumber,
\end{eqnarray}
leading to the mean field Hamiltonian

\begin{eqnarray}
H_{MF} =
E_{0}+\lambda\sum_{i\sigma}\tilde{b}^{\dag}_{i\sigma}\tilde{b}_{i\sigma}
- 2\sum_{\langle ij \rangle} J_{ij} A_{ij}
(\tilde{A}^{\dag}_{ij}+\tilde{A}_{ij}),
\nonumber
\end{eqnarray}
where we have taken  $A_{ij}=\langle \tilde{A}_{ij}\rangle$ to be real.

First consider the case that all the  bonds are identical by symmetry, and
assuming no spontaneous dimerization, then all $A_{ij}$ must be the same
$A_{ij}=A$, .  In this case $E_{0}=\frac{1}{2}NzJS^{2} - 2\lambda NS +
JA^{2}Nz$, where $z$ is the coordiantion number.  $H_{MF}$ can be
diagonalized by going to momentum space and perfroming the Bogoliubov
transformation:

\begin{eqnarray}
H_{MF}=E_{0}-\lambda N
+\sum_{\bf{k}}\omega_{\bf{k}}(\alpha^{\dag}_{\bf{k}}\alpha_{\bf{k}} +
\beta^{\dag}_{\bf{k}}\beta_{\bf{k}} + 1), \nonumber
\end{eqnarray}
where $\omega_{\bf{k}} = [\lambda^{2} -
(J\tilde{A}z\gamma_{\bf{k}})^{2}]^{\frac{1}{2}}$, $\gamma_{\bf{k}} =
\frac{1}{z} \sum_ \delta e^{i\bf{k} \cdot \delta} = \sum_{i=1}^{d} \cos
k_{i}/d$. At $T=0$, the energy should be minimized with respect to
$\lambda$ and $A$, yielding the set of self-consistent equations:

\begin{eqnarray}
S + \frac{1}{2} &=& \frac{1}{2N}\sum_{\bf{k}}\frac{\mu} {(\mu^{2}-
\gamma_{\bf{k}}^{2})^{\frac{1}{2}}}, \nonumber\\
\tilde{A} &=& \frac{1}{2N}\sum_{\bf{k}}\frac{\gamma^{2}_{\bf{k}}} {(\mu^{2}-
\gamma_{\bf{k}}^{2})^{\frac{1}{2}}}, \nonumber
\end{eqnarray}
where we define $\mu\equiv\lambda/(J\tilde{A}z)$.  An essential point of the
theory is that a non-zero mean field amplitude $A$, which gives rise to
boson hopping, indicates short-ranged antiferromagnetic order.  Long-ranged
order is achieved if the hopping amplitude is sufficiently large to give
Bose condensation.  This occurs when these eqs. cannot be satisfied by
having $\mu>1$, in which case $\mu=1$, and the $k=0$ term gives a finite
contribution when converting the momentum sums into integrals:

\begin{eqnarray}
S+\frac{1}{2} &=& m_{s} + \frac{1}{2} \int^{\pi}_{-\pi} \frac{d^{d}{\bf{k}}}
{(2\pi)^{d}} \frac{1} {(1-\gamma_{\bf{k}}^{2})^{\frac{1}{2}}}, \nonumber\\
A &=& m_{s}+\frac{1}{2} \int^{\pi}_{-\pi} \frac{d^{d}{\bf{k}}} {(2\pi)^{d}}
\frac{\gamma_{\bf{k}}^{2}} {(1-\gamma_{\bf{k}}^{2})^{\frac{1}{2}}}, \label{ms2}
\end{eqnarray}

It has been shown that the condensate density $m_{s}$ is also the
sublattice magnetization. For the Heisenberg antiferromagnet on a square
lattice, it was found that Bose condensation occurs for all $S>S_{c}$,
where $S_{c}=0.2$, with a gapless linear excitation spectrum characteristic
of spin waves.  For $S<S_{c}$, $\mu>1$, and there is an energy gap for
excitations.  Thus, for all physical values of $S$, there is
AFLRO.

On the other hand, if two such planes are coupled together
antiferromagnetically, and the interplane coupling is very large compared
to intraplane coupling, the ground state is obviously a valence bond solid
of interplane singlets, and the intraplane correlation length is zero.
Thus, there must be at least one phase transition as the interplane
coupling is increased.  We now analyze this for general $S$ using Schwinger
boson MFT.

The Hamiltonian in this case is

\begin{eqnarray}
H=J\sum_{\langle ij \rangle} {\bf{S}}_{i} {\bf{S}}_{j} +
J_{\perp}\sum_{\langle ij
\rangle_{z}} {\bf{S}}_{i} {\bf{S}}_{j}, \nonumber
\end{eqnarray}
where $\sum_{\langle ij \rangle}$ sums over n.n. on the same plane and
$\sum_{\langle ij \rangle _{z}}$ sums over n.n. on different planes.
Since there is still translational invariance, the mean field Lagrange
multiplier will be the same on all sites.  However, the lack of symmetry
between intraplane and interplane bonds means two mean field ampitudes must
be introduced for the bond decoupling.  Letting these be $A$ and $B$
respectively, and taking them to be both real, the mean field Hamiltonian
is now

\begin{eqnarray}
H_{MF} = &&E_{0} + \lambda\sum_{i\sigma} (\tilde{b}^{\dag}_{i\sigma}
\tilde{b}_{i\sigma}) - 2JA\sum_{\langle ij \rangle}
(\tilde{A}^{\dag}_{ij}+\tilde{A}_{ij})
\nonumber\\
&-& 2J_{\perp}B\sum_{\langle ij \rangle _{z}}
(\tilde{A}^{\dag}_{ij}+\tilde{A}_{ij})\nonumber,
\end{eqnarray}
where $E_{0} = 2NJS^{2}-2\lambda NS + 4JA^{2}N + NJ_{\perp}S^{2}/2 +
J_{\perp}B^{2}N$.  As before, we diagonalize $H_{MF}$ through the
Bogoliubov transformation, giving

\begin{eqnarray}
H_{MF}=E_{0}-\lambda N
+\sum_{\bf{k}\sigma}\omega_{\bf{k}\sigma}(\alpha^{\dag}_{\bf{k}\sigma}\alpha
_{\bf{k}\sigma} + \beta^{\dag}_{\bf{k}\sigma}\beta_{\bf{k}\sigma} + 1),
\nonumber
\end{eqnarray}
The excitation energies are given by

\begin{eqnarray}
\omega_{{\bf{k}},\sigma} = [\lambda^{2} - (2JA\sum_{i=1}^{d} \cos k_{i} +
J_{\perp}B\sigma)^{2}] ^{\frac{1}{2}}. \nonumber
\end{eqnarray}
where $\sigma= \pm 1$.  Minimizing $H_{MF}$ with respect to $\lambda$, $A$, and
$B$ gives the self-consistent equations:

\begin{eqnarray}
S+\frac{1}{2} &=& \frac{1}{2N} \sum_{\bf{k},\alpha} \frac{\mu} {(\mu^{2} -
\Gamma_{\bf{k},\alpha}^{2})^{\frac{1}{2}}} \nonumber\\
A &=& \frac{1}{2N} \sum_{\bf{k},\alpha} \frac{\Gamma_{\bf{k},\alpha}}
{(\mu^{2} - \Gamma_{\bf{k},\alpha}^{2})^{\frac{1}{2}}}
(\frac{\sum_{i=1}^{d} \cos k_{i}} {d})\nonumber\\
B &=& \frac{1}{2N} \sum_{\bf{k},\alpha} \frac{\Gamma_{\bf{k},\alpha}
\alpha} {(\mu^{2} - \Gamma_{\bf{k},\alpha}^{2})^{\frac{1}{2}}},
\label{consist1}
\end{eqnarray}
where $\Gamma_{\bf{k},\alpha} = (\sum_{i=1}^{d} \cos k_{i} + Q \alpha)/d$ and
$Q=J_{\perp}B/(2JA)$.  Note that $\mu$, the excitation gap,
must be greater than or equal $1+Q/d$. In particular, in the case of bose
condensation, the value of $\mu$ is fixed to $1+Q/d$ and hence the
summations in Eq. (\ref{consist1}) turn out to be a function of the
parameter $Q$ only.  The magnetization $m_{s}$ is calculated by solving the
self-consistent equations with the summations converted into intergrals:

\begin{eqnarray}
S+\frac{1}{2} &=& m_{s} + \frac{1}{4} \int \frac{d^{d} {\bf{k}}} {(2\pi)^{d}}
\sum_{\alpha} \frac{\mu_{0}} {(\mu_{0}^{2} -
\Gamma_{\bf{k},\alpha}^{2})^{\frac{1}{2}}} \nonumber\\
\tilde{A} &=& m_{s}+\frac{1}{4} \int \frac{d^{d} {\bf{k}}} {(2\pi)^{d}}
\sum_{\alpha} \frac{\Gamma_{\bf{k},\alpha}} {(\mu_{0}^{2} -
\Gamma_{\bf{k},\alpha}^{2})^{\frac{1}{2}}} (\frac{\sum_{i=1}^{d} \cos
k_{i}} {d} )\nonumber\\
\tilde{B} &=& m_{s}+\frac{1}{4} \int \frac{d^{d} {\bf{k}}} {(2\pi)^{d}}
\sum_{\alpha} \frac{\Gamma_{\bf{k},\alpha} \alpha} {(\mu_{0}^{2} -
\Gamma_{\bf{k},\alpha}^{2})^{\frac{1}{2}}}, \label{consist2}
\end{eqnarray}
where $\mu_{0} = 1+Q/d$.  These equations \cite{Millis} hold so long as they
give
$m_{s}>0$, ie., Bose condensation, otherwise Eqs. (\ref{consist2}) should
be used with $\mu$ also as an unknown parameter.  In principle, we can
solve for $Q$ and $m_{s}$ or $\mu$.  In practice, the form of these
equations allow us to avoid this by plugging in an arbitrary values of $Q$
into the equations to find out $m_{s}$ or $\mu$, and then $A$ and $B$.  The
self-consistency is then reduced to using the values of  $A$, $B$, and $Q$
to determine the
value of $\beta \equiv J_{\perp}/J$.

The behavior of $Q$ as a function of $\beta$ for $S=1/2$ is shown in Fig.
\ref{qvsj}, and is representative of all $S$.  Discounting the trivial
solutions
$Q=0$ and $Q=\infty$, corresponding to independent planes and IVBS
respectively, there are $Q\neq 0$ solutions indicating both intraplane and
interplane correlations.  For small  $\beta$, there is only one solution,
with Q increasing from $0$ with $\beta$.  For $\beta > 4(S+1/2)^{2}$, a
second branch of solution, beginning at infinity appears.  The two
solutions merge at some larger value of $\beta$, and beyond that, only the
trivial solutions remain.  The significance of these solutions can be
understood if we consider the energy $E(Q)$ obtained by minimizing the
energy with respect to all other parameters except $Q$.  Then, the
non-trivial solutions are extrema of $E(Q)$.  Thus, for $\beta <
\beta_{0}=4(S+1/2)^{2}$, the solution of $Q$ corresponds to a global
minimum in $E(Q)$, and describes the ground state.  For $\beta >
\beta_{0}$, the upper branch corresponds to a local maximum while the lower
branch remains a local minimum.  The local maximum begins at $Q=\infty$ at
$\beta_{0}$, and moves towards the local minimum with increasing $\beta$.
Eventually, the two extrema merge into a saddle point at $\beta_{2}$.
Beyond that, $E(Q)$ is  strictly decreasing with $Q$.  By continuity, this
means somewhere between $\beta_{0}$ and $\beta_{2}$, $E(\infty)$ must cross
from being greater than $E(Q_{1})$ to less than it, where $Q_{1}$ is the
lower branch solution. Thus, at this value $\beta_{1}$, the ground state
jumps from the 2D correlated state described by $Q_{1}$ to the interplane
VBS state.

We can solve for the value of $m_{s}$ at the non-trivial solutions.
Initially, $m_{s}$ increases with increasing $Q$, but eventually will
decrease, vanishing at some $Q_{c}$. For sufficiently large $S$, $Q_{c}$
will belong to the upper branch (maximum energy) solution.  More
importantly,  in the case where $Q_{c}$ lies in the lower branch, its
$\beta$ value changes from less than to greater than $\beta_{1}$ with
increasing $S$. Thus, the transition from the LRO'ed state to disordered
state is second order for small $S$, but becomes first order for larger
$S$.  In the former case, there is a subsequent transition from a
disordered state with finite $Q$, hence with both interplane and intraplane
short-ranged correlations, to the $Q=\infty$ state with only interplane
correlations. Along with the jump in $Q$ is a discontinous jump in the gap.
It is tempting to associate this jump as a transition from some disordered
state associate with a single plane to the non-degenerate IVBS. More likely
this transition is probably an artifact of the Schwinger boson MFT, and
indicates a relative sharp drop in the intraplane correlation length and a
sharp rise in the gap. This is similar to the finite temperature MFT solution
for a single plane, where $A$, hence short-ranged correlation, drops to zero
above some finite temperatures \cite{Sarker}. In the latter case of first order
transition in
sublattice magnetization, the ground state jumps from one with LRO to the
IVBS state.  Since this latter state should be the correct ground state
only in the $\beta$ goes to infinity limit, we interpret this as the MFT
way  of showing a transition into a disordered state with a very short
intraplane correlation length.  The behavior of $m_{s}$ and the gap
$\Delta$ as a function of $\beta$ is shown in Fig. \ref{fig2} for
representative
values of $S$.  Fig. 2c shows an example of reentrance, where LRO first
develops with increasing $\beta$, but is subsequently destroyed when
$\beta$ gets too large.  This occurs for $S$ smaller than approximately
0.2, the MFT value of $S$ below which the ground state has no LRO, but
greater than approximately 0.13, the minimum value of $S$ for LRO at some
$\beta$.

The phase diagram of $S$ vs. $\beta$ is shown in Fig. \ref{phase}.  For $S <
0.13$,
the ground state is always disordered.  For $0.13<S<0.2$, the system
undergoes first a disorder-order and then a order-disorder continuous
transition with increasing $\beta$.  For $S>0.2$, there is LRO for $\beta =
0$, and only the order-disorder transition remains.  This transition is
continuous until it terminates at a tricritical point at $S \approx 0.35,
\beta \approx 2.92$, beyond which the continuous transition is preempted by a
first order transition.  Thus, for $S>0.35$, there are values of $\beta$ where
the LRO'ed state is not the ground state, but is nevertheless metastable.
The continuous transition phase boundary remains metastable until $S \approx
0.4$, beyond which the $m_{s}=0_{+}$ state moves into the upper branch and
becomes unstable.  In all cases of $S$ where a disordered ground state with
finite $Q$ exists, a subsequent "first-order transition" occurs, with a
discontinuous jump in $Q$ and the gap $\Delta$. As mentioned above, we
interpret the jump as unphysical, and represents in reality a relatively
sharp drop in the 2D correlation length.

We can understand why large $S$ favors a first order transition quite
simply in terms of spin wave theory \cite{Hida1,Hida2}.  The Neel state energy
is $E_{N} =
S^{2}(2Jz+J_{\perp})$ while the energy of the IVBS state is
$E_{V}=J_{\perp}S(S+1)$.  Equating the two implies an estimate for the
first order transition at $\beta_{1}$ of the order of $S$ for large $S$.
Within spin wave theory, the sublattice magnetization is given by $m_{s}$
in Eq. (\ref{consist2}) with $B/A=1$.  For large $\beta$, the integral on
the LHS scales as $\sqrt{\beta}$.  If we set $m_{s}=0$ as an estimate for
the critical value $\beta_{c}$ for continuous transition, then $\beta_{c}$
is of order $S^{2}$.  Thus, for large $S$, $\beta_{1}$ is much less than
$\beta_{c}$.

Within MFT, the tricritical point and even the metastable continuous
transition boundary occurs below the minimum physical value of $S=1/2$.
Thus, a first order transition is predicted for all physical systems
described by the model.
In fact, the sublattice magnetization jump at transition for $S=1/2$ is
about 30\% of that at $\beta=0$, clearly contradicting the results of
numerical work on the model for $S=1/2$, which supports a continuous
transition in the same universality class as the finite temperature
transition of the $3D$ classical Heisenberg model.  On the other hand,
there is no reason to expect the Schwinger boson MFT to give the exact
answer, so the true position of the tricritical point might very well be
above $S=1/2$.  This is particularly so since by relaxing the local
constraint to a global one, unphysical states are included in the mean
field solution, and the MF energy is not even variational.  Thus, using
these MF energies to find the position of first order transition is
necessarily suspect.  Nevertheless, we believe the prediction of larger $S$
favoring a first order transition to be correct, and the nature of phase
transition in the bilayer system is non-universal.  For example, the
transition for $S=1/2$ may become first order if there is a sufficiently
large next nearest neighbor ferromagnetic interaction.  Conversely, first
order transition may become continuous if frustration is introduced.  In
other words, the value of $S$ at the tricritical point can be changed by
enlarging the parameter space.  The seeming contradiction to the fact that
the 2+1 $D$ NLSM has only continuous transition is resolved by noting that
the mapping of the Heisenberg model into the NLSM is legitimate only if the
correlation length is long, which does not have to be the case of the
disordered state close to a first order transition.  Also of interest is
that with sufficient frustration, the single-layered system can be
disordered for $S=1/2$ or other physical values, and the reentrance
behavior for small $S$ discussed above in the bilayered system can be
physically observed.

Within our MFT and according to general arguments, first order transition
implies the existence of metastable states with finite sublattice
magnetization.  This may lead to observeable dynamics characteristic of
macroscopic quantum tunneling.  It would also be of significance with
respect to Monte Carlo type numerical calculations
\cite{Hida3,Dagotto,Sandvik1,Sandvik2} due to problems of being
"stuck" in the metastable minimum.  For example, the first order transition
may be missed if the metastability persists till the would-be continuous
transition.

\section{Extension to the Fractional Dimensions}

We have seen that  for the 2D bilayered square lattice  antiferromagnet,
Schwinger boson MFT shows the physically interesting case of $S=1/2$  as
undergoing  a first order transition.  Since the continuous transition is
favored by small $S$, hence increasing quantum fluctuations, one way of
getting  a continuous  transition in MFT for $S=1/2$ is to go to a
dimension below $2$.  In this section, we show that indeed this is the
case. En route, we present the phase diagram of $S$ vs. $d$ (Fig.
\ref{critdim}) for a
single layer. These non-integer dimension results may be relevant to the
physics of the Heisenberg antiferromagnet on percolating clusters, which
have fractal dimensionality.

 We first perform the Schwinger boson MFT  for a single hypercubic lattice
in $d$ dimension (square lattice for $d=2$).  The self-consistent equations
(Eq. (\ref{ms2})) depend on $d$ only through the momentum sum, which must
be analytically continued to non-integer dimensions.  This can be done by
using the gaussian identity:

\begin{eqnarray}
\frac{1} {(1 \pm \gamma_{\bf{k}})^{\frac{1}{2}}} = 2 \sqrt{\frac{d}{\pi}}
\int^{\infty}_{0} dx ~e^{-x^{2}(d \pm \sum^{d}_{i=1} \cos k_{i})},
\label{gauss}
\end{eqnarray}
to rewrite Eq. (\ref{ms2}) as

\begin{eqnarray}
\frac{2d}{\pi} \int^{\infty}_{0} dxdy~ e^{-(x^{2}+y^{2})d} (\int^{\pi}_{-\pi}
\frac{dk}{2\pi} e^{-\cos k (x^{2}-y^{2})} ) ^{d}. \label{bessel}
\end{eqnarray}
In this form, the analytic continuation to arbitrary $d$ is obvious (see
Appendix).  In
Fig. \ref{msgap}, the result for $S=1/2$ is shown.  As $d$ is decreased below
$2$,
$m_{s}$ decreases and vanishes at some critical dimension $d_{c}=1.46$.
Below $d_{c}$, the excitation spectrum has a gap, which rises with
decreasing $d$.  These behaviors are representative for all $S$. However,
it is known that for the simple Heisenberg Hamiltonian, $1/2$ integer spin
chains and integer spin chains are intrinsically different in that the
former should be gapless,
which can be understood as due to the presence of a topological term the
appropriate NLSM.  Thus MFT must break down for $1/2$ integer spins even
qualitatively as $d$ gets sufficiently close to $1$,  and $\Delta$  must
decrease again.

Next we generalize our bilayer calculation to two coupled hypercubes in $d$
dimension. The analytic continuation of Eqs (\ref{consist2}) can again be done
using the gaussian identity (\ref{gauss}). As expected,  a continuous
transition can now be observed within MFT if $d$ is reduced sufficiently from
$2$. For $d <1.46$, the critical dimension for LRO for a single hypercube, the
reentrance
seen with increasing $\beta$ for $S<0.2$ is seen for $S=1/2$.  So far, we
have concentrated on lowering the dimension from $2$.  Of course, raising
it would have the opposite effect.  For example, for two $3D$ hypercubes,
the $S=1/2$ transition would be strongly first order within MFT, and so
even taking into account inaccuracy of MFT, strongly implies a first order
transition.

Finally we discuss the critical phenomena of the continuous transition of this
model. Analyzing Eqs. (2) and (3) close to the transition, we find the
staggered magnetization vanishes linearly, while the gap vanishes as $(\beta -
\beta_{c})^{s}$ with $s=1/(d-1)$ for $d<3$, and $s=1/2$ for $d>3$ (there are
logarithm corrections at $d=3$). These MF exponents are the same as those for
the finite temperature transition of a single hypercube with the substitution
$d\rightarrow d+1$, reflecting the quantum nature of the present transition.

\section{Appendix}

Notice that in Eq. (\ref{bessel}), the integral inside the bracket is a
modified Bessel function of the first kind $I_{0}(x^{2}-y^{2})$. Therefore the
first equation of Eq. (\ref{consist2}) can be written as:

\begin{eqnarray}
S+\frac{1}{2} = m_{s} + \frac{2d}{\pi} \int^{\infty}_{0} dxdy
(e^{-(x^{2}+y^{2})} I_{0}(x^{2}-y^{2}))^{d}. \nonumber
\end{eqnarray}
Take the transformation:

\begin{eqnarray}
u=x^{2}-y^{2}, ~~~~~~v=x^{2}+y^{2} \nonumber
\end{eqnarray}
with $v\geq |u|$, for the integration variable will give us:

\begin{eqnarray}
S+\frac{1}{2} &=& m_{s} + \frac{2d}{\pi} \int^{\infty}_{-\infty} du
(I_{0}(u))^{d} \int^{\infty}_{|u|} dv \frac{e^{-vd}}{\sqrt{v^{2}-u^{2}}}
\nonumber \\
&=& m_{s} \frac{2d}{\pi} \int^{\infty}_{0} du (I_{0}(u))^{d}
K_{0}(ud).\nonumber
\end{eqnarray}
It reduces the final formula into a single integral of $I_{0}$ and $K_{0}$,
modified Bessel function of the second kind. In this form, the integral indeed
converges much faster than the original form in Eq. (\ref{consist2}) and
consequently save much of the computation time. The same trick is also applied
to both interplane and intraplane mean field equations.

\newpage
\begin{figure}
\caption{The behavior of $Q$ as a function of $\beta$ for $S=1/2$. It is
representative of all $S$. $\beta_{0} = 4 $ and $\beta_{2} \approx 4.36$. The
star symbol stands for the location of 1st order transition in which $\beta
\approx 4.25$.}
\label{qvsj}
\end{figure}

\begin{figure}
\caption{Magnetization (solid line) $m_{s}$ and spin gap (dashed line) $\Delta
/ \beta$ as a function of $\beta$. (a) shows 1st order transition with the
absent of 2nd order transition for $S=1/2$ while both transitions are observed
for smaller $S$ in (b). (c), reentrance of magnetization occurs for $S < 0.2$.}
\label{fig2}
\end{figure}

\begin{figure}
\caption{Phase diagram of $S$ vs $\beta$ for $d=2$. 1st order transiton exists
as long as $\beta < 2.92$ in which 1st order transition coincides with 2nd
order transition (star symbol). The zero magnetization line (solid) terminates
at a tricritical point, $\beta \approx 3.38$, beyond which the zero
magnetization state is no longer stable.}
\label{phase}
\end{figure}

\begin{figure}
\caption{Phase diagram of $S$ vs $d$ for a single layer. The curve goes to
infinity as $d$ tends to 1.}
\label{critdim}
\end{figure}

\begin{figure}
\caption{Magnetization $m_{s}$ and spin gap $\Delta$ for $1\leq d \leq 2$ and
$S=1/2$. The critical dimension that separates order and disorder state is
$\approx 1.46$.}
\label{msgap}
\end{figure}



\begin{references}
\bibitem{Arovas} D. P. Arovas and A. Auerbach, Phys. Rev. B {\bf 38}, 316
(1988).
\bibitem{Sarker} S. Sarker, C. Jayaprakash, H. R. Krishnamurthy and M. Ma, {\bf
40}, 5028 (1989).
\bibitem{Millis} A. J. Millis and H. Monien, Phys. Rev. B {\bf 50}, 16606
(1994).
\bibitem{Hida1} T. Matsuda and K. Hida, J. Phys. Soc. Jpn. {\bf 59}, 2223
(1990).
\bibitem{Hida2} K. Hida, J. Phys. Soc. Jpn. {\bf 59}, 2230 (1990).
\bibitem{Hida3} K. Hida, J. Phys. Soc. Jpn. {\bf 61}, 1013 (1992).
\bibitem{Dagotto} E. Dagotto, J. Riera, D. Scalapino, Phys. Rev. B {\bf 45},
5744 (1992).
\bibitem{Sandvik1} A. W. Sandvik and D. J. Scalapino, Phys. Rev. Lett. {\bf
72}, 2777 (1994).
\bibitem{Sandvik2} A. W. Sandvik and M. Veki$\acute{c}$, J. Low Temp. Phys.
{\bf 99}, 367 (1995).
\bibitem{Yoshioka} D. Yoshioka et. al., to be published.
\bibitem{Maekawa} S. Maekawa et. al., to be published.
\bibitem{Haldane} F. D. M. Haldane, Phys. Rev. Lett. {\bf 61}, 1029 (1988).
\end{references}
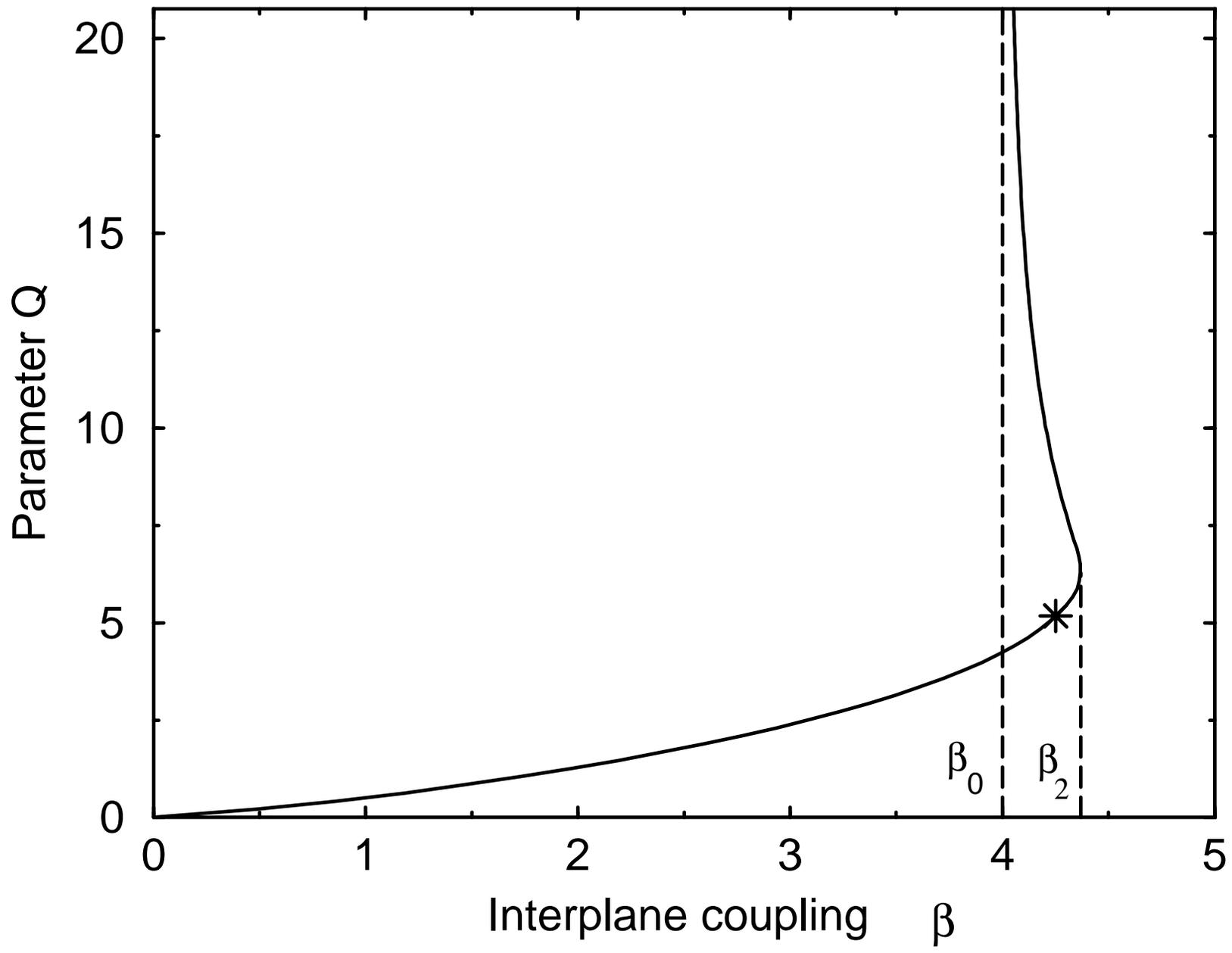
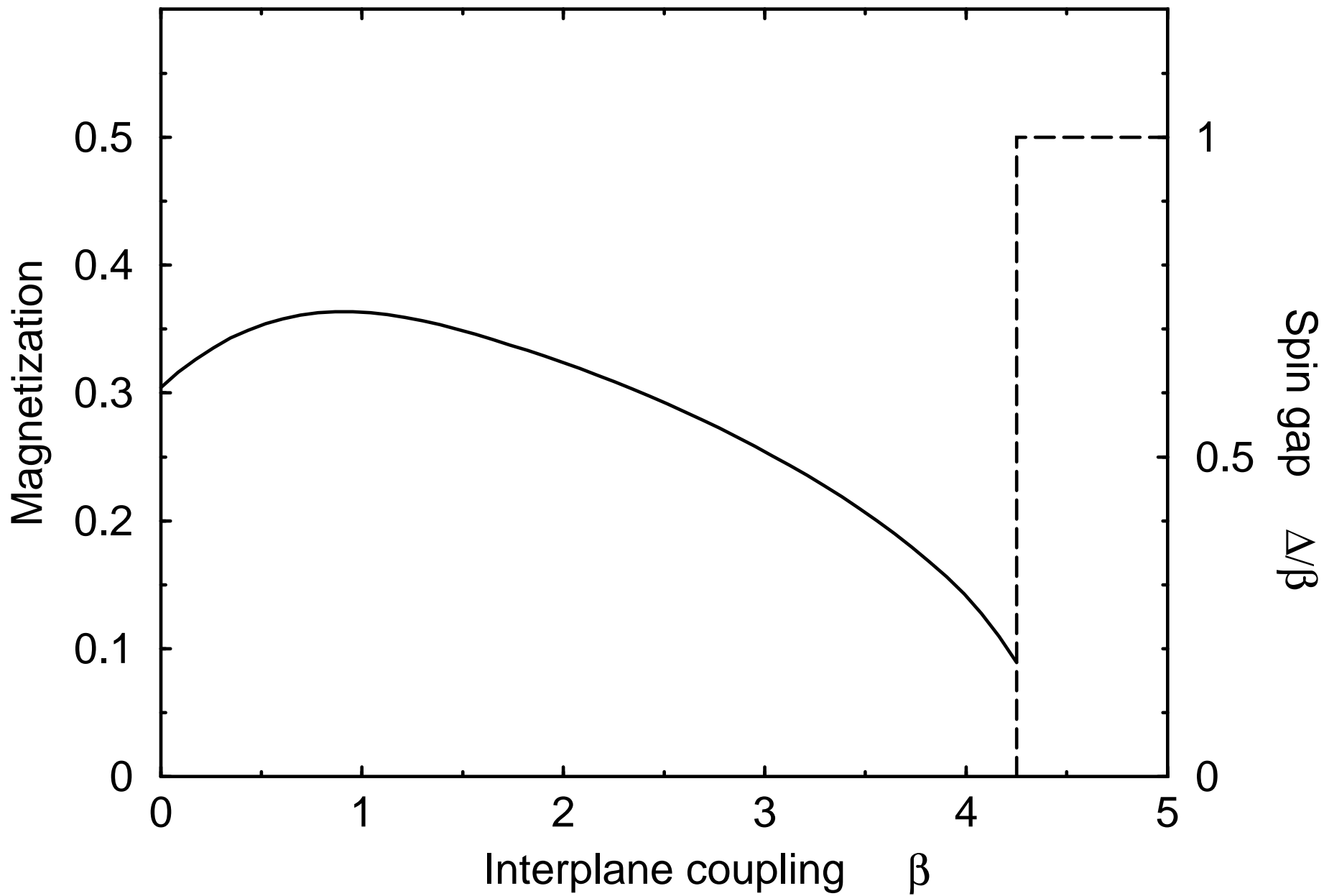
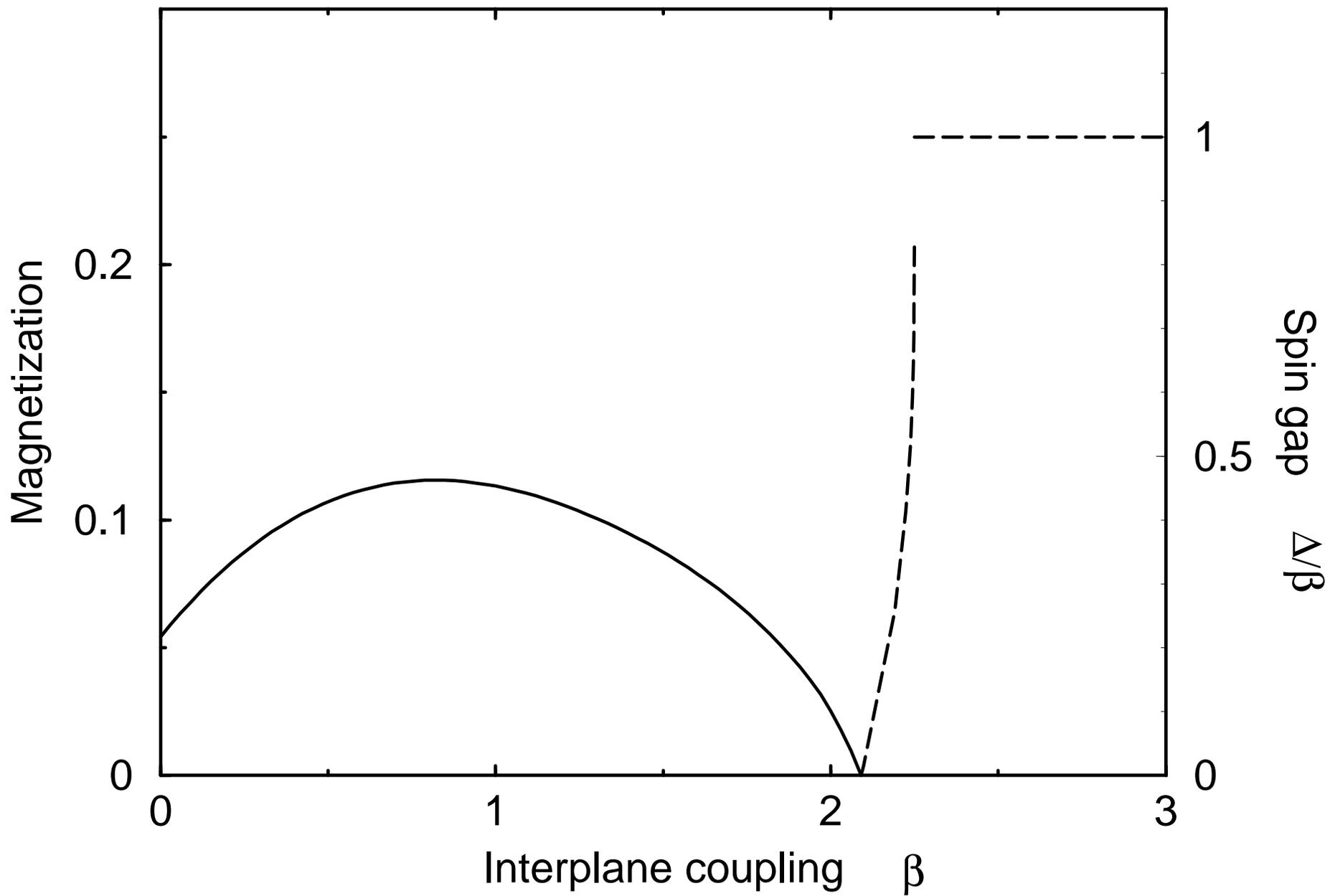
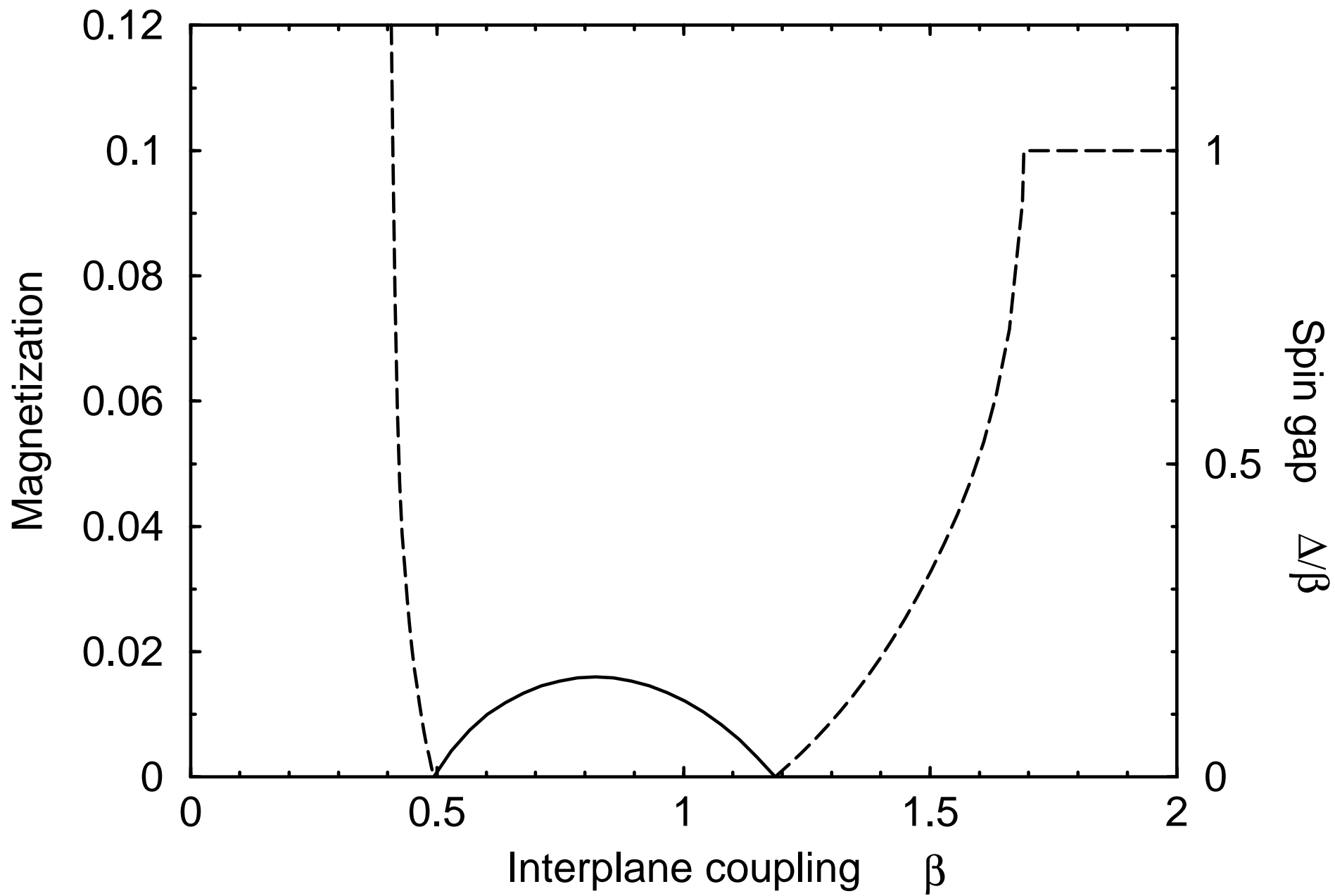
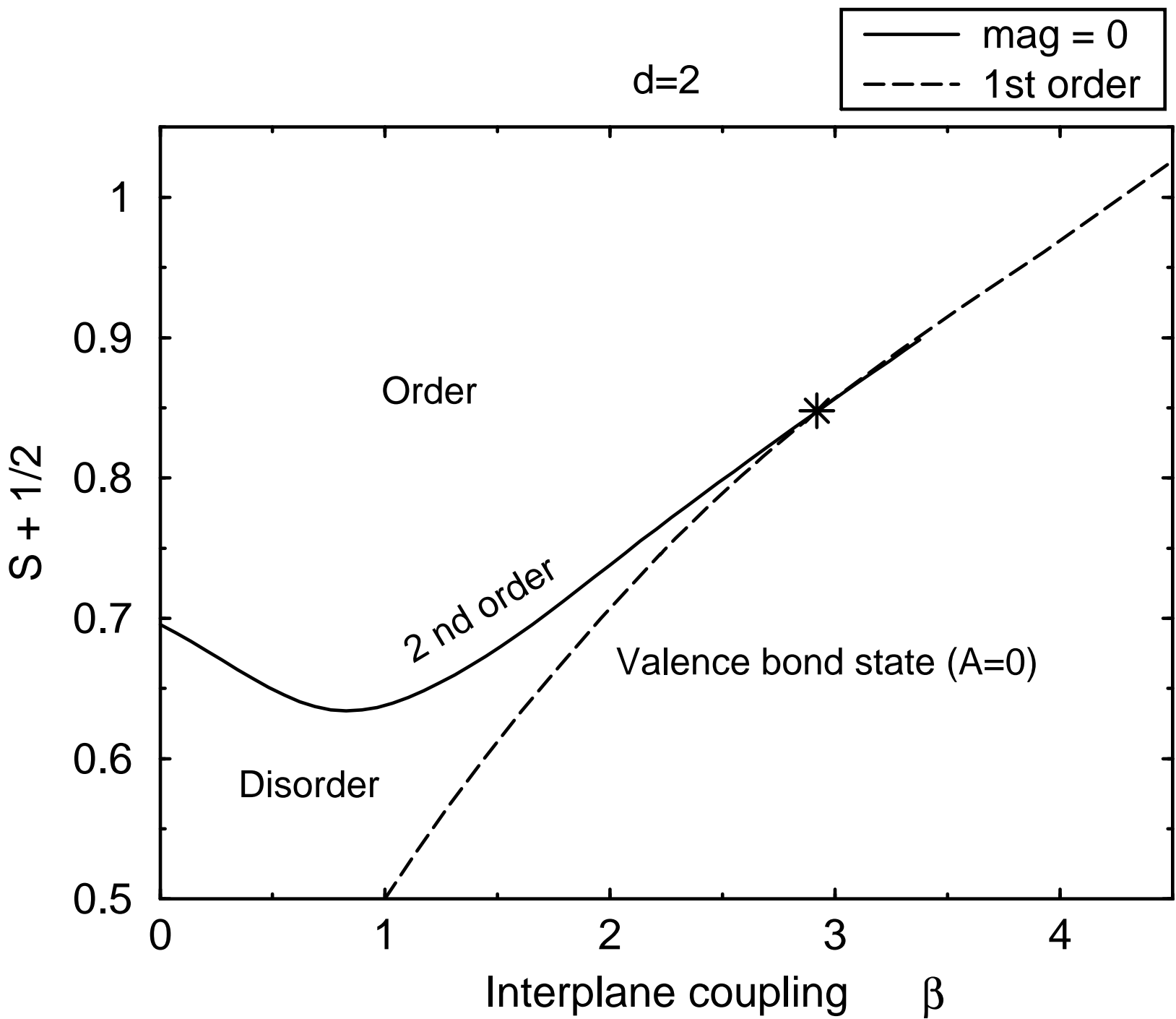
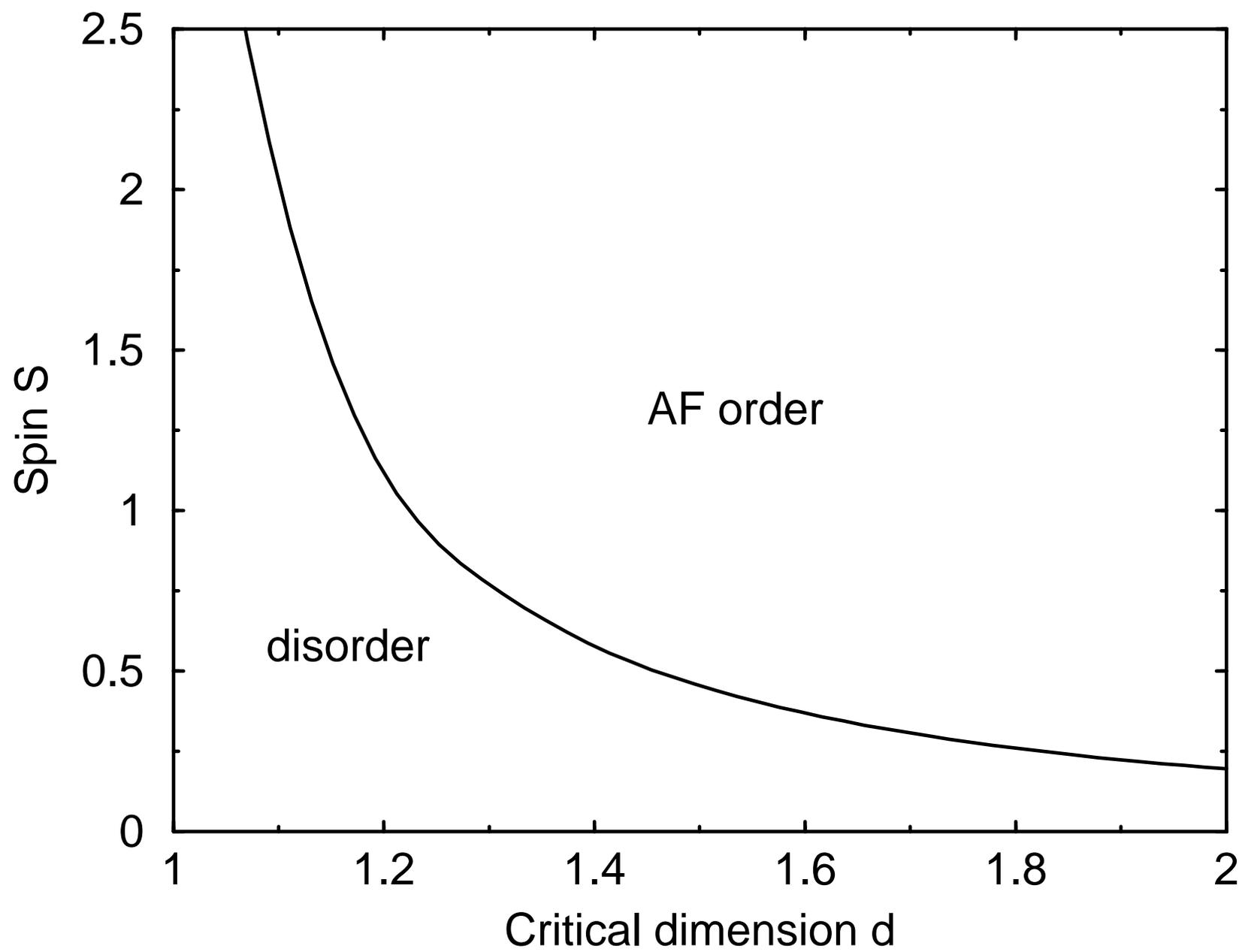
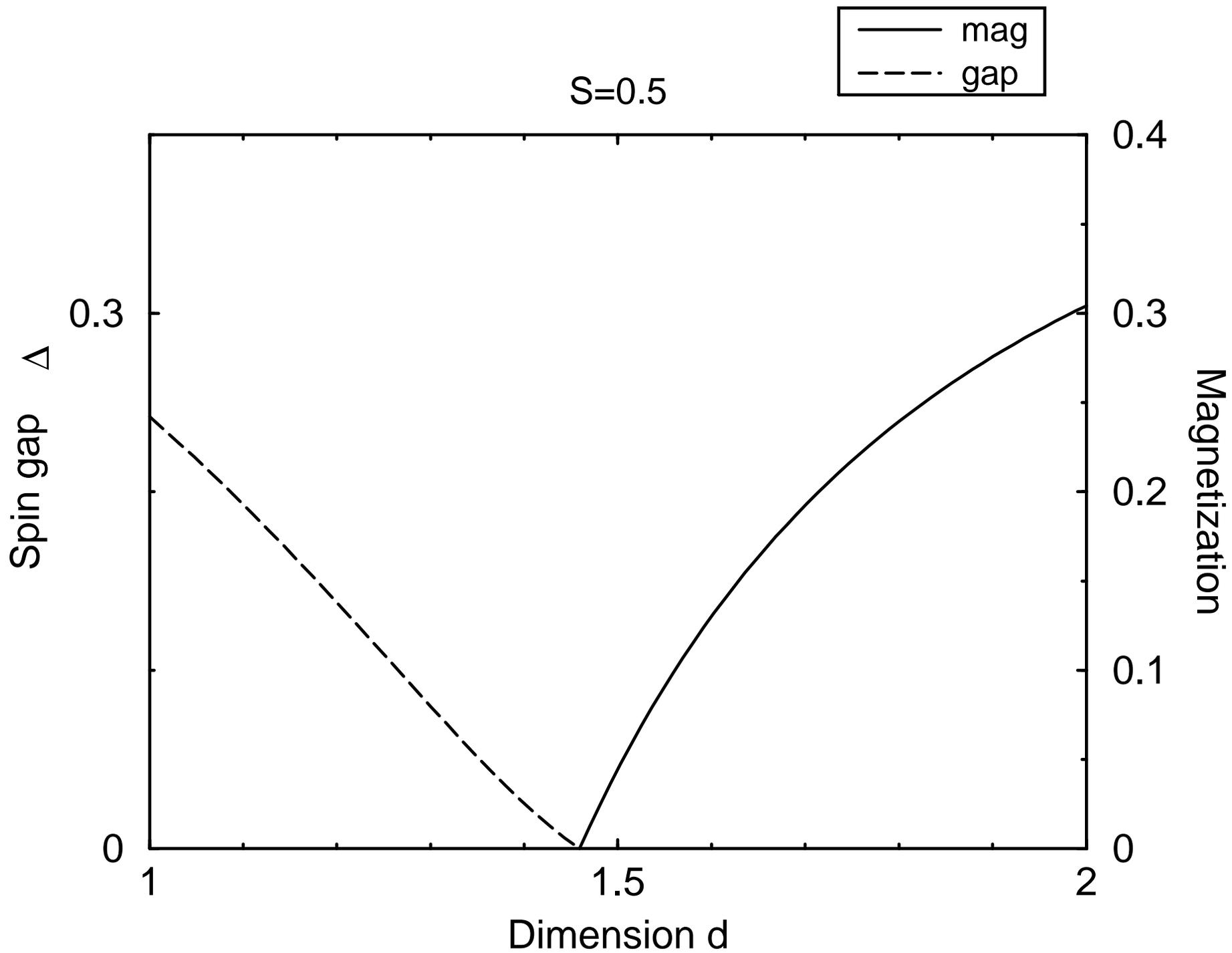
\end{document}